\documentclass[
    ,final            
  ]
  {aipproc}

\layoutstyle{6x9}

\newcommand\chandra{{\it Chandra}}
\newcommand\eg{{\it e.g.}}
\newcommand\etal{{\it et al.}}
\newcommand\ie{{\it i.e.}}
\newcommand\cf{{\it cf.}}
\newcommand\etc{{\it etc}}
\newcommand\vs{{\it vs}}
\newcommand\ergps{{\rm\ erg\ s^{-1}}}
\newcommand\lxcool{L_{\rm X,cool}}

\begin{document}

\title{Radio Mode Outbursts in Giant Elliptical Galaxies}

\classification{95.85.Nv,98.52.Eh,98.58.Kh,98.62.Nx}
\keywords      {galaxies:elliptical, intergalactic medium, cooling
  flows, galaxies:jets} 

\author{Paul Nulsen}{
  address={Harvard-Smithsonian Center for Astrophysics, 60 Garden St,
    Cambridge, MA 02138, USA}
}

\author{Christine Jones}{
  address={Harvard-Smithsonian Center for Astrophysics, 60 Garden St,
    Cambridge, MA 02138, USA}
}

\author{William Forman}{
  address={Harvard-Smithsonian Center for Astrophysics, 60 Garden St,
    Cambridge, MA 02138, USA}
}

\author{Eugene Churazov}{
  address={Max-Planck-Institut f\"ur Astrophysik,
    Karl-Schwarzschild-Strasse 1, 85741 Garching, Germany}
  ,altaddress={Space Research Institute (IKI), Profsoyuznaya 84/32,
    Moscow 117810, Russia}
}

\author{Brian McNamara}{
  address={Department of Physics and Astronomy, University of
    Waterloo, 200 University Avenue West, Waterloo, Ontario, N2L 3G1,
    Canada}
  ,altaddress={Perimeter Institute for Theoretical Physics, 31
    Caroline St. N., Waterloo, Ontario, N2L 2Y5, Canada} 
}

\author{Laurence David}{
  address={Harvard-Smithsonian Center for Astrophysics, 60 Garden St,
    Cambridge, MA 02138, USA}
}

\author{Stephen Murray}{
  address={Harvard-Smithsonian Center for Astrophysics, 60 Garden St,
    Cambridge, MA 02138, USA}
}

\begin{abstract}
  Outbursts from active galactic nuclei (AGN) affect the hot
  atmospheres of isolated giant elliptical galaxies (gE's), as well as
  those in groups and clusters of galaxies.  \chandra{} observations of
  a sample of nearby gE's show that the average power of AGN outbursts
  is sufficient to stop their hot atmospheres from cooling and forming
  stars, consistent with radio mode feedback models.  The outbursts are
  intermittent, with duty cycles that increases with size.
\end{abstract}

\maketitle


\section{Introduction}

\chandra{} X-ray observations have shown that energies deposited by
AGN outbursts at the centers of clusters and groups of galaxies are
about sufficient to prevent the hot gas there from cooling, consistent
with models in which ``radio mode'' AGN feedback limits cooling and
star formation \citep[\eg,][]{brm04,df06,rmn06,mn07}.
There is similar evidence of AGN feedback in galaxy groups and
isolated elliptical galaxies \citep[\eg,][]{fj01,jfv02,mnj06}.
The lack of outbursts in some clusters with short central cooling
times \citep[\eg,][]{wmm04} implies that outbursts are intermittent.
Moderately strong shocks \citep[\eg,][]{swa02,mnw05,mnj06}
require large, sustained increases in AGN power, also supporting
intermittency. 

The long timescales of AGN outbursts ($>$ Myr) permit us only a single
snapshot of them.  Thus, to study the cycles of intermittent AGN
outbursts, we must resort to samples.  Here, we present results of
\chandra{} observations of a sample of nearby giant elliptical
galaxies (gE's).

\section{Outbursts in the elliptical galaxy sample}

Based on the samples of Beuing \etal{} \citep{bdb99} and O'Sullivan
\etal{} \citep{opc03}, Jones \etal{} (in prep.) have assembled a
sample of $\simeq160$ nearby gE's with $L_K > 10^{10}\rm \ L_\odot$
that have been observed with \chandra.  After removal of point
sources, including a component due to unresolved point sources
\citep{rcs07}, 104 of these galaxies show emission from diffuse hot
gas.  They constitute our sample of gE's with significant hot
atmospheres, and 24 of them show cavities due to AGN outbursts.

Many \chandra{} exposures used here are shallow, so that some
outbursts are likely to have been missed.  In some known radio
sources, the detected X-ray cavities are much smaller than the radio
source, likely causing their outburst energies to be underestimated.
Nor is the \chandra{} archive a well controlled, complete sample.  We
estimate that roughly $1/3$ of nearby gE's have been observed with
\chandra{}.  However, many nearby gE's that were detected by {\it
  ROSAT} have been observed with \chandra{}.  Thus, while the
sample is incomplete, it is probably representative of nearby, X-ray
bright gE's.

\begin{figure}
  \includegraphics[height=.3\textheight,angle=270]{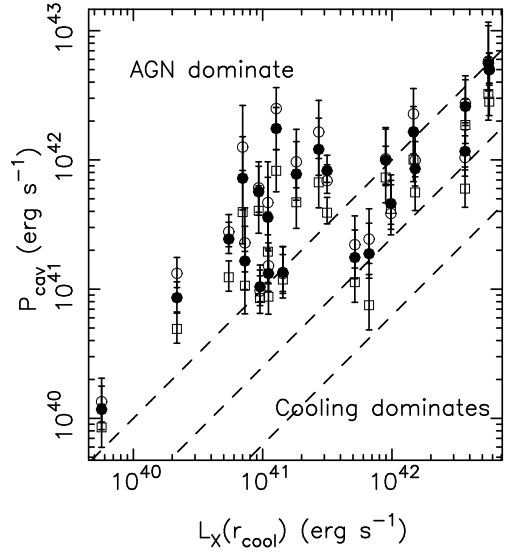}
  \caption{Cavity power \vs{} cooling power for outbursts in the gE
    sample.  Filled markers, open circles and open squares show
    $P_{\rm sonic}$, $P_{\rm buoy}$ and $P_{\rm refill}$,
    respectively, with 90\% confidence ranges.  Dashed lines show
    $P_{\rm cav} = \lxcool$ for energy inputs of $1pV$, $4pV$ and
    $16pV$ per cavity, respectively, top to bottom.} \label{fig:gebirzan}
\end{figure}

Outburst (cavity) powers were determined as in \citep{brm04,rmn06}.
The energy of each cavity is taken to be the product of its pressure,
$p$, and its volume, $V$ (\ie, $1pV$ of energy per cavity).  Three age
estimates are used, the sound crossing time, $t_{\rm sonic}$, the
buoyant crossing time, $t_{\rm buoy}$ and the refill time, $t_{\rm
  refill}$, providing a guide to the systematic uncertainty in the age
estimates \citep{brm04}.  These give three estimates of the mean
power, $pV/t$, for an outburst, $P_{\rm sonic}$, $P_{\rm buoy}$ and
$P_{\rm refill}$.  The cooling power is determined for each atmosphere
as the power it radiates from within the region where the gas cooling
time is shorter than 7.7 Gyr (look back time to $z = 1$).
Cavity power estimates are plotted against cooling powers in
Fig.~\ref{fig:gebirzan}.  Dashed lines show where cavity power equals
cooling power for energy inputs of $1pV$, $4pV$ and $16pV$ per cavity.
At $4pV$ per cavity, $P_{\rm sonic} > \lxcool$ for every outburst in
the sample.  In marked contrast, outburst powers scatter about cooling
powers for rich clusters \citep{rmn06}.

Outbursts are detected in $\sim 1/4$ of the gE sample \citep[\cf{}
  $\sim 2/3$ of rich clusters,][]{df06}.  To study the impact of
intermittent outbursts on gE's, we assume that our sample is
representative of randomly selected stages of the AGN outburst cycle.
Totals of the cavity power estimates are $P_{\rm tot,sonic} = 2.6
\times 10^{43} \ergps$, $P_{\rm tot,buoy} = 2.9 \times 10^{43} \ergps$
and $P_{\rm tot,refill} = 1.5 \times 10^{43} \ergps$, while the total
cooling power for all 104 gE's is $L_{\rm tot,cool} = 8.7 \times
10^{43} \ergps$.  Thus, the ratio of the average cooling power to the
average outburst power for the sample is $\langle \lxcool \rangle /
\langle P_{\rm cav} \rangle = 3.4$, 3.0 and 5.6, for the three power
estimates, respectively.

\begin{figure}
  \includegraphics[height=.4\textheight,angle=270]{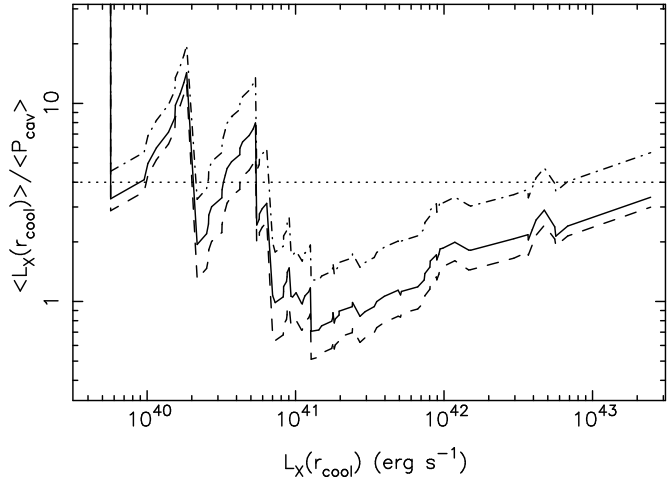}
  \caption{Ratio of cumulative average cooling power to cumulative
    average outburst power \vs{} cooling power.  The full, dashed and
    dot-dashed lines show results for $P_{\rm sonic}$, $P_{\rm buoy}$
    and $P_{\rm refill}$, respectively.} \label{fig:cumav}
\end{figure}

The minimum energy needed to make a cavity is its thermal energy plus
the work required to inflate it, \ie, its enthalpy.  For a cavity
(radio lobe) dominated by relativistic gas, this is $4pV$.  Additional
energy is lost driving shocks or sound waves \citep{swa02,mnw05}
adiabatic losses, leakage of cosmic rays, \etc.  Since cavity powers
are also likely to be underestimated, the means of $\langle \lxcool
\rangle / \langle P_{\rm cav} \rangle$ above (for $1pV$ per cavity)
show that an energy input of $4pV$ per cavity is probably sufficient
for the time averaged outburst power to match the cooling power for
this sample.  Thus, AGN feedback can regulate cooling and star
formation in these lower mass systems, as well as in clusters.

As powerful systems dominate these averages, we apply a further check
for the fainter gE's.  After sorting by $\lxcool$, ratios of the
cumulative cooling power to the cumulative outburst powers are plotted
in Fig.~\ref{fig:cumav}.  At small cooling powers, small numbers (and
intermittency) make the results very noisy.  Nevertheless, the average
ratio is consistent with a value of $\sim4$ across the whole sample.
Thus, intermittent AGN outbursts that deposit $4pV$ per cavity can
regulate cooling in all gE's with significant hot atmospheres.

\begin{figure}
  \includegraphics[height=.4\textheight,angle=270]{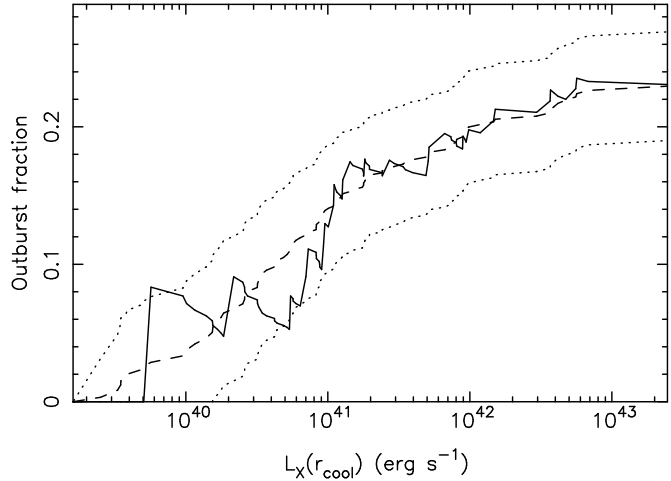}
  \caption{Cumulative outburst fraction \vs{} cooling power for the
    sample (full line).  The expected value for the best fitting model
    for this sample (dashed) and its one sigma range (dotted) are
    also shown.} \label{fig:duty}
\end{figure}

\section{Outburst duty cycle}

The cumulative fraction of gE's with outbursts, plotted against
$\lxcool$ in Fig.~\ref{fig:duty}, increases with $\lxcool$.  If
outburst probability (\ie, duty cycle), $p(\lxcool)$, is linear in
$\log (\lxcool)$, a maximum likelihood fit gives $p = 0.09$ at
$\lxcool = 10^{40} \ergps$ and $p = 0.49$ at $10^{43} \ergps$.  The
fit is just inconsistent with a constant duty cycle at the 90\%
confidence level.  Extrapolating to $\lxcool \simeq 10^{44.5} \ergps$
gives $p \simeq 0.7$, consistent with results for rich clusters
\citep{df06}.  At 0.23, the overall fraction of outbursts in our
sample is also inconsistent with rich clusters, unless the duty cycle
increases with $\lxcool$.  Lastly, the smaller duty cycle in
smaller systems requires outbursts to have larger powers relative to
$\lxcool$ if average outburst power is to match the cooling power.
This is seen in Fig.~\ref{fig:gebirzan}, where the trend of the data
points is flatter than the dashed lines.

In summary, AGN outbursts can limit cooling in the entire range of
gE's with significant hot atmospheres.  The outbursts are
intermittent, with duty cycles that increase with cooling luminosity.


\begin{theacknowledgments}
  PEJN was supported by NASA grant NAS8-03060.
\end{theacknowledgments}



\bibliographystyle{aipproc}   

\newcommand\apj{ApJ}
\newcommand\araa{ARAA}
\newcommand\apjl{ApJ}
\newcommand\mnras{MNRAS}
\newcommand\aap{AA}
\newcommand\nat{Nature}

\bibliography{nulsen}


\end{document}